\documentclass[preprint,12pt]{elsarticle}

\usepackage{amssymb}
\usepackage{amsmath}
\usepackage{lineno}

\usepackage{algorithm}
\usepackage[noend]{algpseudocode}

\usepackage{hyperref}
\hypersetup{
  colorlinks,
  citecolor=Violet,
  linkcolor=Red,
  urlcolor=Blue}
\biboptions{numbers,comma,square,sort&compress}

\usepackage{nicefrac}

\usepackage{mathtools}
\usepackage{siunitx}
\DeclarePairedDelimiterXPP\BigOSI[2]%
  {\mathcal{O}}{(}{)}{}%
  {\SI{#1}{#2}}

\newcommand{\revision}[1]{\textcolor[rgb]{0.0,0.0,0.0}{#1}}

\journal{Scripta Materialia}

\begin{document}

\begin{frontmatter}

%% Title, authors and addresses

%% use the tnoteref command within \title for footnotes;
%% use the tnotetext command for theassociated footnote;
%% use the fnref command within \author or \address for footnotes;
%% use the fntext command for theassociated footnote;
%% use the corref command within \author for corresponding author footnotes;
%% use the cortext command for theassociated footnote;
%% use the ead command for the email address,
%% and the form \ead[url] for the home page:
%% \title{Title\tnoteref{label1}}
%% \tnotetext[label1]{}
%% \author{Name\corref{cor1}\fnref{label2}}
%% \ead{email address}
%% \ead[url]{home page}
%% \fntext[label2]{}
%% \cortext[cor1]{}
%% \affiliation{organization={},
%%             addressline={},
%%             city={},
%%             postcode={},
%%             state={},
%%             country={}}
%% \fntext[label3]{}

\title{Accelerating the discovery of low-energy structure configurations: a computational approach that integrates first-principles calculations, Monte Carlo sampling, and Machine Learning}

%% use optional labels to link authors explicitly to addresses:
%% \author[label1,label2]{}
%% \affiliation[label1]{organization={},
%%             addressline={},
%%             city={},
%%             postcode={},
%%             state={},
%%             country={}}
%%
%% \affiliation[label2]{organization={},
%%             addressline={},
%%             city={},
%%             postcode={},
%%             state={},
%%             country={}}

\author[inst1]{Md Rajib Khan Musa}

\affiliation[inst1]{organization={Department of Mechanical Engineering, Villanova University},%Department and Organization
            %addressline={ Villanova University}, 
            city={Villanova},
            postcode={19085}, 
            state={PA},
            country={USA}}

\author[inst1]{Yichen Qian}

\author[inst1]{Jie Peng}

\author[inst1]{David Cereceda\corref{cor1}}
\cortext[cor1]{Corresponding author}
\ead{david.cereceda@villanova.edu}

\begin{abstract}
Finding Minimum Energy Configurations (MECs) is essential in fields such as physics, chemistry, and materials science, as they represent the most stable states of the systems. In particular, identifying such MECs in multi-component alloys considered candidate PFMs is key because it determines the most stable arrangement of atoms within the alloy, directly influencing its phase stability, structural integrity, and thermo-mechanical properties. However, since the search space grows exponentially with the number of atoms considered, obtaining such MECs using computationally expensive first-principles DFT calculations often results in a cumbersome task. 
To escape the above compromise between physical fidelity and computational efficiency, we have developed a novel physics-based data-driven approach that combines Monte Carlo sampling, first-principles DFT calculations, and Machine Learning to accelerate the discovery of MECs in multi-component alloys.
More specifically, we have leveraged well-established Cluster Expansion (CE) techniques with Local Outlier Factor models to establish strategies that enhance the reliability of the CE method.  
In this work, we demonstrated the capabilities of the proposed approach for the particular case of a tungsten-based quaternary high-entropy alloy. However, the method is applicable to other types of alloys and enables a wide range of applications.
\end{abstract}

%%Graphical abstract
% \begin{graphicalabstract}
% \includegraphics{grabs}
% \end{graphicalabstract}

%%Research highlights
% \begin{highlights}
% \item Research highlight 1
% \item Research highlight 2
% \end{highlights}

\begin{keyword}
%% keywords here, in the form: keyword \sep keyword
Monte-Carlo  \sep DFT \sep cluster expansion \sep high entropy alloy \sep machine learning \sep data-driven Monte-Carlo
%% PACS codes here, in the form: \PACS code \sep code
%\PACS 0000 \sep 1111
%% MSC codes here, in the form: \MSC code \sep code
%% or \MSC[2008] code \sep code (2000 is the default)
%\MSC 0000 \sep 1111
\end{keyword}

\end{frontmatter}

%\linenumbers

%% main text

%%%-------------------------------------------------
%%%-------------------------\input{1_Introduction}
%%%-------------------------------------------------

%\section{Introduction}\label{sec_Intro}
%\newpage
%--------------------------------------------------------
%--------- HEAs and the importance of SRO

High-entropy alloys (HEAs), first conceptualized in the early 2000s as a blend of five or more elements with individual concentrations between 5 and 35 atom percent \cite{yeh2004nanostructured}, have gained attention as a promising class of materials due to their exceptional properties \cite{tsai2014high,li2016metastable,george2019high,murty2019high,han2024multifunctional}.
Recent works have indicated that the distribution of elements in these multi-principal element materials is not entirely random, particularly at lower temperatures where local ordering phenomena, such as short-range order (SRO) and local clustering, emerge to minimize the free energy of the system \cite{li2019strengthening,ding2018tunable,zhang2020short,chen2021direct,hsiao2022data}.
Indeed, the SRO is found to play a crucial role in determining the stability of the crystal structure \cite{ding2018tunable,smith2024competition}. Recent studies have also reported its significant impact as a driver of properties, including stacking fault energy \cite{ding2018tunable}, tensile strength \cite{miracle2017critical,han2024ubiquitous}, resistance to fatigue \cite{zhang2021improving}, magnetization \cite{feng2017effects,walsh2021magnetically}, work-hardening \cite{zhang2020short}, corrosion resistance \cite{liu2018effect,xie2021percolation,scully2020controlling,blades2024tuning}, and irradiation damage \cite{zhang2023effect,lu2017radiation,cao2021does}, among others.

%--------------------------------------------------------
%--------- Previous experimental and computational works for SRO. Need to find MECs for accurate SRO predictions

Previous experimental studies have investigated the local structural characteristics of multi-principal element alloys (MPEAs) using extended X-ray absorption fine structure (EXAFS) techniques \cite{zhang2017local,joress2023exafs}. However, the quantitative analysis of SRO in MPEAs using EXAFS faces significant challenges. These include, for example, the large number of parameters introduced by these alloys and the poor elemental contrast due to the presence of elements with similar atomic numbers \cite{joress2023exafs}.
%
%Quantifying SRO using computational techniques offers the possibility to overcome such challenges, but the accuracy of their calculations strongly depend on the physical fidelity of the underlying atomistic model.  %the arrangement of the different chemical species within the model.
Quantifying SRO using computational techniques can help address these challenges, but the accuracy of their calculations heavily depends on the physical fidelity of the underlying atomistic model.
Identifying the minimum energy configurations (MECs) that represent the most stable arrangement of atoms within the alloy is, therefore, critical as a preliminary step in calculating SRO. 
%

%--------------------------------------------------------
%---------  Traditional MC-DFT 

Earlier computational works have successfully coupled density functional theory (DFT) calculations with Monte-Carlo (MC) sampling in an approach (denoted as MC-DFT hereinafter) that allows for the discovery of such MECs \cite{hastings1970monte,tamm2015atomic,kang2023density,zhang2017local,feng2017effects,ding2018tunable,zhang2020short,samanta2024machine,sheriff2024quantifying,cao2024capturing,yichen_mcdft_2024}.
In essence, this MC-DFT method samples the phase space to find MECs by randomly swapping the chemical elements between atom locations, and the probability of accepting or rejecting a swapped configuration depends on its energy and the Monte Carlo temperature.
%
%--------------------------------------------------------
%---------  Vastness of configuration space, required number of swaps and computational cost of DFT calculations
%
%Despite the suitability of the MC-DFT approach in finding MECs, large cells, a significant number of initial independent samples, and numerous MC swaps are required to converge to a realistic atomistic configuration 
While the MC-DFT approach is suitable for identifying MECs, it requires large cells, a substantial number of initial independent samples, and numerous MC swaps to converge to a realistic atomistic configuration
\cite{ding2018tunable,samanta2024machine,sheriff2024quantifying}.
For example, performing 20-25 swaps trials per atom in a $4\times4\times4$ \textit{bcc} structure containing 128 atoms will need approximately between 2500-3200 MC steps.  
Evaluating the energy of each swapped configuration using highly accurate yet computationally expensive first-principles DFT calculations becomes a cumbersome task. 
And this process becomes even more challenging when considering larger systems due to the vastness of their configuration space.

%--------------------------------------------------------
%--------- This work

% To address these bottlenecks
To overcome the above bottlenecks, we propose here a novel computational approach that integrates MC sampling, first-principles DFT calculations, and machine learning techniques to accelerate the discovery of MECs.
%
%It is grounded in the principles of the Cluster Expansion (CE) method \cite{Kikuchi1951, Sanchez1984, Zunger2002, DeFontaine1994} and it also incorporates a novel Local Outlier Factor (LOF) model \cite{}, enabling the rapid and accurate prediction of the energy of the swapped configurations in the traditional MC-DFT approach. 
It is grounded in the principles of the Cluster Expansion (CE) method \cite{Kikuchi1951, Sanchez1984, Zunger2002, DeFontaine1994} and incorporates a Local Outlier Factor (LOF) model \cite{breunig2000lof}, which facilitates the rapid and accurate prediction of the energy of the swapped configurations, enhancing the computational efficiency of the traditional MC-DFT framework.

%--------------------------------------------------------
%--------- Target alloy system: HEA1

The target alloy system chosen to demonstrate the capabilities of our proposed approach is a tungsten-based quaternary and equiatomic HEA, WCrTiTa \cite{sobieraj2020chemical}.
We selected this tungsten-based HEA as this class of materials, designed to withstand the extreme environments \cite{el2019outstanding, waseem2017powder}, has shown superior mechanical properties at high temperatures, a superior melting point (above 2873 K), enhanced radiation resistance to heavy ion irradiation, and negligible radiation hardening when compared to pure tungsten \cite{waseem2017powder,zou2015ultrastrong,senkov2018development,el2021helium,el2019outstanding,senkov2011mechanical, qian2021using, qian2023ab}, one of the leading candidates for plasma-facing materials (PFMs) in fusion energy applications.
However, the proposed method is applicable to other types of alloys and enables a wide range of applications.

%%%-------------------------------------------------
%%%-------------------------\input{2_Methods}
%%%-------------------------------------------------

%\section{Computational Methods}\label{sec_Methods}

\begin{figure}[hbtp!]
    \includegraphics[width=\textwidth]{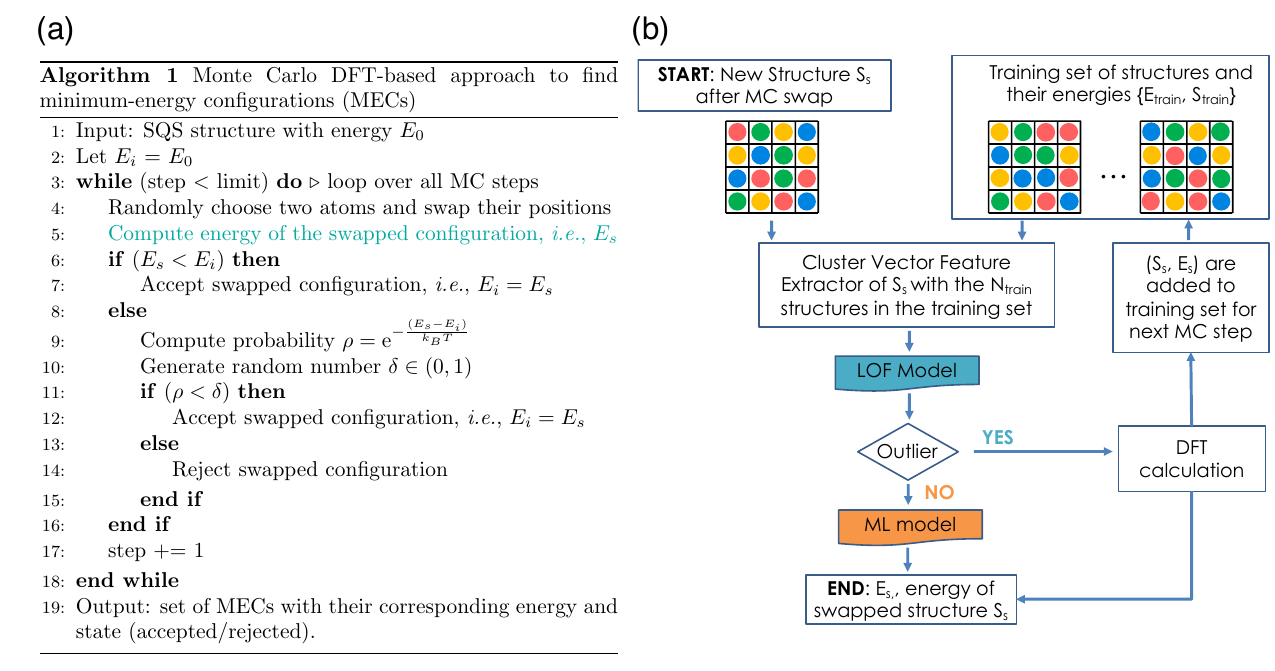}
	\caption{\label{fig_workflow} (a) Detailed description of the MC-DFT algorithm. Highlighted is the step where the accelerated MC-DFT (aMC-DFT) is applied; (b) detailed flowchart of the aMC-DFT for calculating the energy of the swapped configuration, a loop that is repeated every MC step.}
\end{figure}

Figure~\ref{fig_workflow} shows the details of the proposed accelerated MC-DFT  algorithm (referred to as a-MCDFT hereinafter) and its framework for evaluating the energy of a given atomistic structure.
The main difference between the traditional MC-DFT and the a-MCDFT methods lies in how they calculate the energy of the swapped configurations. % generated in every MC step.
In the MC-DFT method, these energy evaluations are always performed using DFT calculations in every MC step.
% Conversely / in contrast
%  using a less computationally-expensive yet accurate surrogate model
%In contrast, the a-MCDFT approach allows for the energy of swapped configurations to be calculated using a surrogate model based on the flag provided by a LOF model that compares the newly swapped configuration with those in a training set.
For their part, in the a-MCDFT approach, a LOF module first compares the newly swapped configuration with those in the training set and then, depending on their similarity, calculates the energy of the system using a previously trained surrogate model or DFT calculations.

As stated above, the a-MCDFT framework is based on the principles of the CE method. However, we noticed that the probability of accepting a newly swapped configuration depends not on the energy of the newly swapped configuration $E_s$  but on the energy difference between the most recently accepted and the newly swapped configurations $(E_s - E_i)$. Thus, instead of expressing the free energy as a linear combination of contributions from individual finite-size clusters within the lattice, here we formulate our model using the cluster vector difference and the energy difference between the two structures, i.e.:

%
% \begin{equation}
% \Delta E(\sigma)=\sum_{\alpha} K_{\alpha}\Delta\phi_{\alpha}(\sigma) 
% \label{eq_04}
% \end{equation}

\begin{equation}
%\Delta E(\textnormal{S})=\sum_{\alpha} K_{\alpha}\Delta\phi_{\alpha}(\textnormal{S}) 
\Delta E_{s,j}=\sum_{\alpha} K_{\alpha}\Delta\phi_{\alpha}(\textnormal{S}_s,\textnormal{S}_j) 
%\Delta E(\textnormal{S}_s)=\sum_{\alpha} \sum_{j} K_{\alpha}\Delta\phi_{\alpha}(\textnormal{S}_s,\textnormal{S}_j) 
\label{eq_04}
\end{equation}

% \begin{equation}
% \Delta E(\textnormal{$S_s$})=\sum_{\alpha} K_{\alpha}\Delta\phi_{\alpha}(\textnormal{$S_s$}) 
% \label{eq_04}
% \end{equation}

\noindent where $\Delta \phi_{\alpha}(\textnormal{S}_s,\textnormal{S}_j)$ are the cluster differences for a cluster $\alpha$ between the newly swapped configuration $\textnormal{S}_s$ and the $\textnormal{S}_j$ configuration present in the training set, $\Delta E_{s,j}$ is the energy difference between the two structures, and $K_{\alpha}$ are the parameters obtained when fitting 
the above equation to a reference data set containing ab initio calculations. %or experimental measurements.
%a surrogate model that takes $\Delta \phi_{\alpha}$ as the features and $\Delta E$ as the target.
%
The series is truncated in a certain order, including only clusters up to a certain size or distance between sites. This results in a manageable number of terms in the expansion and allows for a computationally tractable approximation of the properties of the alloy. 

As part of the training process, one also has to minimize the residual $||\mathbf{K}\mathbf{\Delta\phi}-\mathbf{\Delta E}||$. This is a linear least squares problem that can be solved by various methods, such as the normal equations method or singular value decomposition (SVD).
However, it is often the case that the number of clusters included in the expansion, and hence the number of parameters to be determined, is large, and the system of equations is ill-conditioned. Thus, regularization techniques, such as L1 regularization (Lasso) or L2 regularization (Ridge), can be used to obtain a stable and sparse solution for \(\mathbf{K}\).
%
%Interestingly, given a training set containing $N_{train}$ pairs of structures with their energies $\left(E_i, S_i \right)$, obtained exclusively via MC-DFT, there are a total of $N_{train}^2$ pairs of structures available for training this new model. That represents one of the main advantages of our model
%
Interestingly, one of the main advantages of the model formulated above is that the number of data points available for training scales quadratically with the number of DFT calculations performed. %That is, given a training set containing $N_{train}$ pairs of structures with their energies $\left(E_i, S_i \right)$, obtained exclusively via MC-DFT, a total of $\BigOSI{}{N^2}$ pairs of structures and their energies are available for training.

%Given a training set containing $N_{train}$ pairs of structures with their energies $\left(E_i, S_i \right)$ obtained exclusively via MC-DFT, there are a total of $N_{train}^2$ pairs of structures available for training this new model. That represents one of the main advantages of our model: the number of training data points scales quadratically, increasing as $O(N^2)$, with the number of DFT calculations performed.

Once the surrogate model is trained, the energy of a newly swapped configuration, denoted as $S_{s}$, is predicted by calculating the cluster vector differences between the new structure and all training structures, i.e., $\Delta \Phi_{s,j}=\Phi_{s}-\Phi_{j}$ where $j=1,2,3,\ldots, N_{train}$. This yields $N_{train}$ different predictions, $\Delta E_{s,j}$. 
To estimate the energy of the new structure, we could simply sum these predicted energy differences with the corresponding energies of the training structures and calculate their average, i.e.:

\begin{equation}
E_{s}=\frac{1}{N_{train}}\sum_{j=1}^{N_{train}} (E_{j}+ \Delta E_{s,j})
\label{eq_06}
\end{equation}

However, there exists the possibility that for certain predictions, the cluster differences between the new structure and some of the training structures denoted as $\Delta \Phi_{s,j}$, are significantly different from the original training data $\Delta E_{s,j}$. In such cases, the accuracy of the predictions may be compromised. 
To address this issue, we employ a method to identify and exclude ML predictions where $\Delta \Phi_{s,j}$ differs substantially from the original training data.
Such an approach is based on the Local Outlier Factor (LOF) model \cite{breunig2000lof} that can compare new data with the training data and determine whether it is similar (considered an inlier) or an outlier. Using the LOF score obtained from this comparison, we can then adjust the new energy predictions as follows:

\begin{equation}
E_{s}=\frac{1}{N_{train}}\sum_{j=1}^{N_{train}} \alpha_j(E_{j}+ \Delta E_{s,j})
\label{eq_07}
\end{equation}

\begin{equation}
\alpha_j=
\begin{cases}
    0 & \Delta \Phi_{s,j} \ \text{is outlier} \\
    1 & \Delta \Phi_{s,j} \ \text{is inlier}
\end{cases}
\label{eq_03}
\end{equation}

As stated above, the main difference between the traditional MC-DFT and the a-MCDFT methods lies in how they calculate the energy of the swapped configurations. % generated in every MC step.
In the MC-DFT method, these energy evaluations are always performed using DFT calculations in every MC step. In the proposed a-MCDFT method, DFT calculations are only used to predict the energy of the newly swapped configurations when the LOF model flags them as outliers. When a newly swapped is flagged as an inlier, the energy is predicted using a surrogate model, significantly reducing the computational time to perform such calculations. 

All extra DFT calculations performed as part of the a-MCDFT method were done on defect-free $4 \times 4 \times 4$ bcc supercells containing 128 atoms, using the Vienna Ab initio Simulation Package (VASP) \cite{kresse1993ab} with projector augmented wave (PAW) pseudo-potentials \cite{blochl1994projector} and the Perdew-Burke-Ernzerhof (PBE) exchange-correlation functional \cite{PBE}. 
Energy calculations within the MC-DFT method employed a plane wave cutoff energy of 300 eV and a $3 \times 3 \times 3$ \textit{k}-point mesh.
All DFT calculations were performed without spin polarization included as there is no experimental evidence about magnetic properties in the considered W-HEAs containing Cr. Indeed, previous works in the literature have shown that, for the binary W-Cr system, which is important in our study, no magnetic phases are observed from its phase diagram \cite{handbook1992alloy}. 
\revision{The MC temperature used in all our calculations is 100K. The reader referred to our recent work \cite{yichen_mcdft_2024} for more details on how the MC temperature (a hyperparameter within the MC-DFT formulation) can impact the discovery of MECs.} \\

%%%-------------------------------------------------
%%%-------------------------\input{3_Results}
%%%-------------------------------------------------

%\section{Results} \label{sec_Results}

%\subsection{Hyperparameter optimization} \label{sec_results_hyperparameter}
\noindent \emph{Hyperparameter optimization}:
%----This version when not using section headers (Scripta)
%Given the framework described above, it is necessary to find the optimal values for the hyperparameters of the model before predicting the energy of newly swapped configurations.
%----This version when using section headers
As a preliminary step to predict the energy of newly swapped configurations, it is necessary to find the optimal values for the hyperparameters of the model. 
In our case, those hyperparameters are the cutoff radius (pair and triplet) of the cluster vectors, the size of the training set containing $N_{train}$ pairs of structures with their energies $\left(E_i, S_i \right)$ obtained exclusively via MC-DFT, and the number of nearest neighbors needed for the LOF model. 

\begin{figure}[H]
    \centering
    \includegraphics[width=1.0\textwidth]{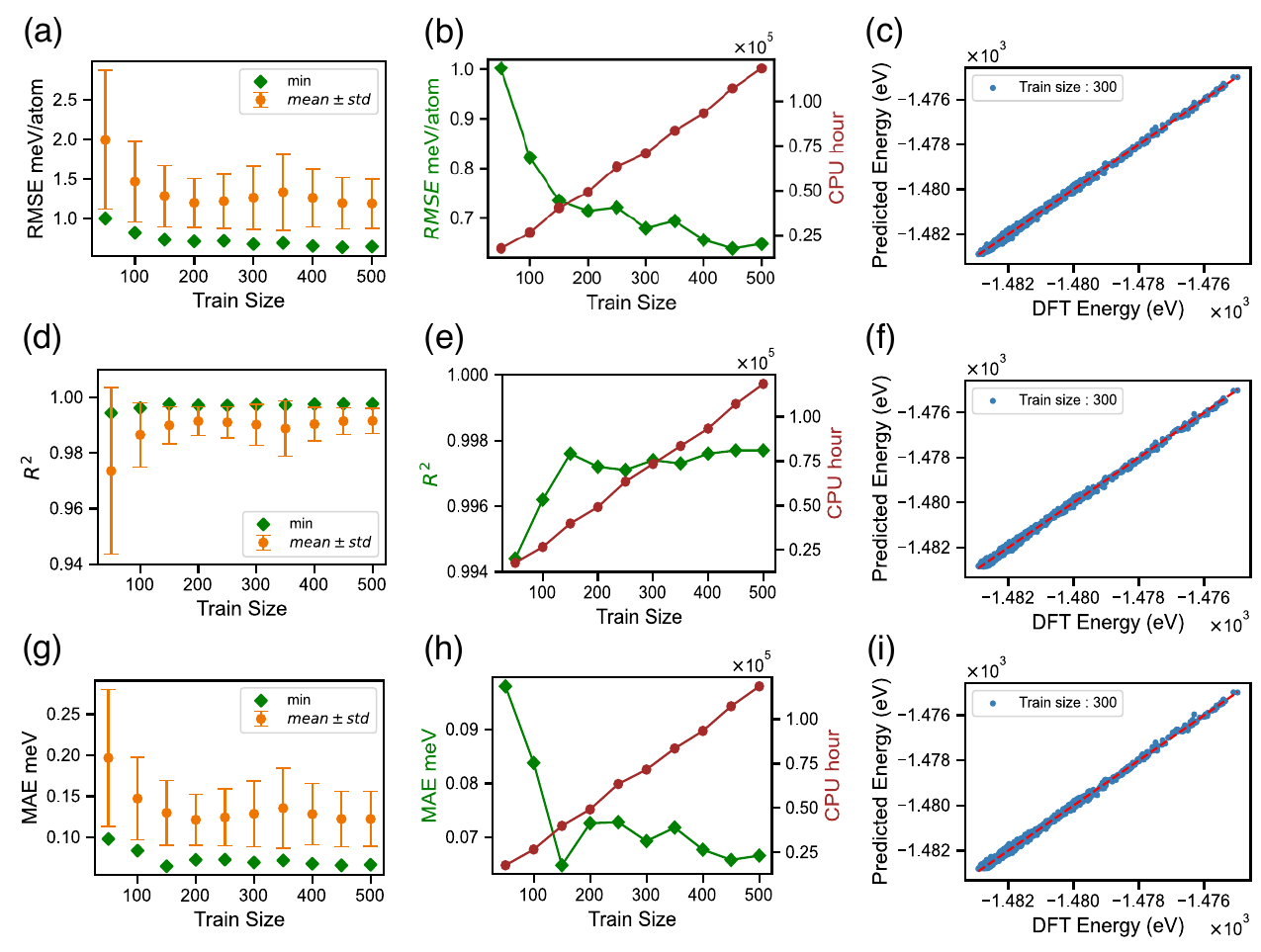}
    \caption{Statistical analysis of the hyperparameter optimization for different training sizes: (a, d, g) Average RMSE, \(R^2\), and MAE values, respectively, for all combinations (orange color) with green diamonds representing the values for the best hyperparameter combinations. (b, e, h) RMSE, \(R^2\), and MAE values, respectively, for the best hyperparameter combinations alongside computational cost (CPU hours). (c, f, i) Predicted vs. DFT energy values produced by a model trained with 300 pairs of structures with their energies $\left(E_i, S_i \right)$ obtained exclusively via MC-DFT, showing the correlation between predicted and actual energy values using the best hyperparameters based on the lowest RMSE, highest \(R^2\), and lowest MAE values, respectively.}
    \label{fig_grid_search}
\end{figure}

For that purpose, we performed a grid search over these four hyperparameters, with the goal of minimizing both the error in the energy predictions and the computational cost. 
The reader is referred to \ref{sect_app_grid_search} for more details about the grid search, whose results are summarized in Fig.~\ref{fig_grid_search}.
These include the statistical analysis based on the following three performance metrics: the Root Mean-Squared Error (RMSE), the Mean Absolute Error (MAE), and the R-squared (R$^2$) error. 
The subfigures in the first column display the performance metrics for all possible combinations within the grid search. They include the mean and standard deviation of all the search space, as well as the configuration with the best value of the performance metric for each training size.  
The second column illustrates the performance metric value for the optimal combination of hyperparameters at each training size, along with the associated computational cost for each case.
As can be seen in these figures, there is a tradeoff between improving the performance metric of interest by increasing the training size and the computational cost of the training process. Thus, we chose a training size of 300 pairs of structures with their energies $\left(E_i, S_i \right)$ obtained exclusively via MC-DFT, a pair cutoff of 8.5 \AA, a triple cutoff of 4.5 \AA, and a No. of neighbors (for the LOF model) of 25, as the values of the four hyperparameters of the surrogate model that will evaluate the energy predictions within the framework of the a-MCDFT method proposed here.
Finally, the third column presents, for the optimal combination of hyperparameters described above, the comparison of the energy of the configurations predicted using the ML model and the actual energies obtained using DFT calculations. \\
%

%%%%%--------------------------------------

\begin{figure}[H]
    \centering
    %%trim={<left> <lower> <right> <upper>}
    \includegraphics[width=1.0\textwidth,trim={2cm 0cm 0cm 2.2cm}]{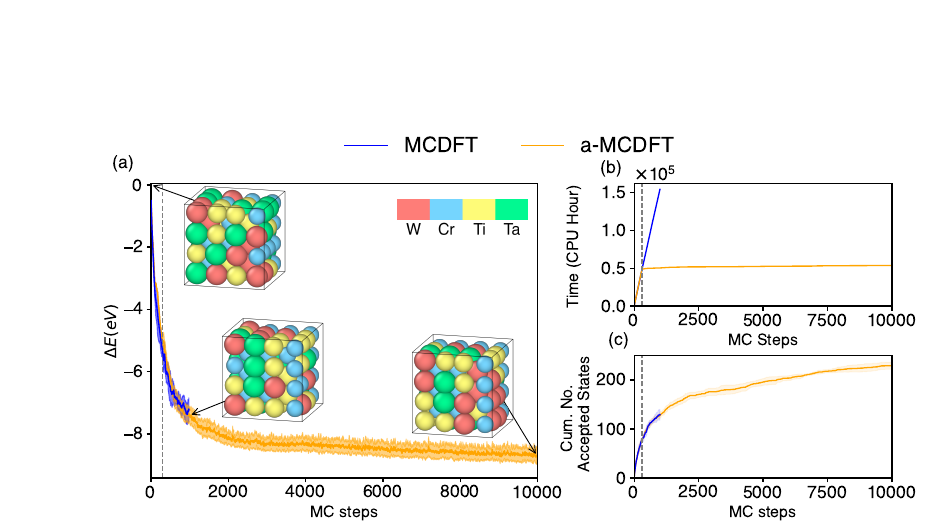}
    \caption{Comparison of minimum energy convergence, computational cost, and acceptance rate among different methods during Monte Carlo (MC) simulations: (a) Energy difference (\(\Delta E\)) as a function of MC steps for MCDFT and a-MCDFT methods. (b) Computational cost (CPU hours) as a function of MC steps for MCDFT and a-MCDFT methods. (c) Cumulative number of accepted states as a function of MC steps for MCDFT and a-MCDFT methods.}
    \label{fig_aMCDFT_predictions}
\end{figure}

%\subsection{Prediction of MECs}\label{results_prediction_MECS}
\noindent \emph{Prediction of MECs}: Once the model has been trained and the optimized set of hyperparameters has been chosen, we applied the a-MCDFT method to find MECs of HEA1, our target alloy system.
We employed a new set of MC-DFT data (different from the one used in the grid search) to train the surrogate model of the a-MCDFT algorithm during the first 300 MC steps. 
\revision{The reader is referred to \ref{sect_app_ML_training} for more details about the training of the surrogate model.}
Then, for each MC step, whenever a newly swapped configuration was generated, the workflow described in Fig.~\ref{fig_workflow} was executed: a LOF module first compares the newly swapped configuration with those in the training set and, depending on their similarity, calculates the energy of the system using the previously trained surrogate model or DFT calculations. 
We repeated this process three times to obtain several completely independent runs of the predictions of the change in energy with our proposed a-MCDFT method.  
The results are shown in Fig.~\ref{fig_aMCDFT_predictions}, which provides a comparative analysis of the MCDFT and a-MCDFT methods, with a particular focus on their energy convergence, computational cost, and acceptance rate. 

Fig.~\ref{fig_aMCDFT_predictions}a illustrates the energy difference (\(\Delta E\)) with respect to the initial configuration as a function of MC steps. 
The figure reveals several interesting trends that support our motivation to develop this newly a-MCDFT approach. 
Firstly, the energy profile obtained with the a-MCDFT method, from the point where training was halted at 300 MC steps (indicated in the figure with the reference dashed line), closely follows the energy profile obtained with the standard MC-DFT method up to 1000 MC steps, which is the maximum number of MC steps calculated with this baseline method.
Secondly, the a-MCDFT method keeps identifying configurations with lower energies up to $10^4$ MC steps, far beyond the initial 300 MC steps where it stopped its training. 
Such long simulations allow for approximately 78 swap trials per atom, which would be prohibitively expensive using the traditional MC-DFT method alone. 
Thirdly, the high level of accuracy of the a-MCDFT method in predicting the energy of the MECs is confirmed when comparing the energies predicted using the a-MCDFT method and the actual energies obtained using DFT calculations. 
Indeed, as we report in \ref{sect_app_B}, the relative difference between the energies predicted with the a-MCDFT and MC-DFT methods is $\sim 0.022\%$.

Fig.~\ref{fig_aMCDFT_predictions}b shows a comparison of the computational cost of the MC-DFT and the a-MCDFT methods. As can be seen, both approaches exhibit a steep increase in computational cost up to 300 MC steps, where a-MCDFT stopped taking configurations and their energies as training information. 
However, beyond that point, while MC-DFT maintains the same slope given the resource-intensive nature of evaluating the energy of each swapped configuration using DFT calculations, the computational cost of the a-MCDFT approach remains practically flat compared to MC-DFT up to 10$^4$ MC steps, when we stopped the simulations. Indeed, the computational cost of a-MCDFT only increases after stopping its training due to the few configurations identified as outliers by the LOF model (an average of 29 configurations were flagged as outliers between 300 and 10$^4$ MC steps), which require energy evaluations using DFT calculations. In most other cases, where the newly swapped configurations are considered similar to those in the training set, the computational cost of evaluating their energies using the previously trained surrogate model is practically zero.
Furthermore, a comparison of the computational cost and energy levels achieved by both methods after training (the reader is referred to Table~\ref{table_dE_CPU} in \ref{sect_app_B} for more details) reveals the significant improvement that the a-MCDFT represents. 
From 300 to 1000 MC steps, the MC-DFT method achieves a \(dE\) of $-2.408eV$, requiring $1.041\times10^5$ CPU hours to complete those 700 MC steps. That would indicate a convergence speed of $-2.313\times10^{-5}$ eV/CPU hour. 
For their part, the a-MCDFT method reaches a \(dE\) of $-4.146eV$ in just $0.047\times10^5$ CPU hours, which implies a convergence speed of $-88.212\times10^{-5}$ eV/CPU hour. 
The above differences in terms of convergence speed demonstrate that the proposed a-MCDFT can find MECs with lower energies at a fraction of the computational cost that would be needed with the traditional MC-DFT approach.

Finally, Fig.~\ref{fig_aMCDFT_predictions}c illustrates the cumulative number of accepted states through the course of the simulations. This parameter can be used to evaluate the ability of the methods to keep finding low-energy structure configurations. 
Both MCDFT and a-MCDFT follow a similar trend, indicating that a-MCDFT method explores the configuration space similarly to MC-DFT. 
Furthermore, while the slope is more pronounced at the beginning of the simulations, the profiles of the two methods don't seem to reach a saturation point, indicating they both could potentially keep finding structures with lower energies over time. 

\begin{figure}[H]
    \centering
    %%trim={<left> <lower> <right> <upper>}
    \includegraphics[width=1.0\textwidth,trim={2cm 0cm 0cm 2.2cm}]{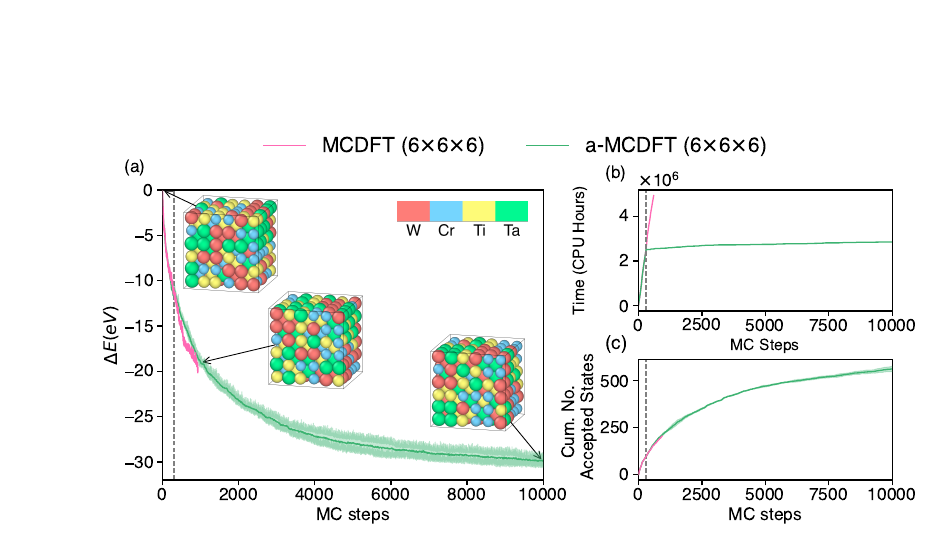}
    \caption{\revision{Scalability of the a-MCDFT method on larger $6 \times 6 \times 6$ supercells, containing 432 atoms. The a-MCDFT simulations on the  $6 \times 6 \times 6$ supercells used the same hyperparameters obtained from the grid search on the $4 \times 4 \times 4$ supercells. During the first 300 MC steps (training size is one of the hyperparameters of the grid search), the surrogate model was re-trained to obtain a new value of the regression coefficients and the regularization parameter. (a) Energy evolution with respect to the energy of the initial SQS structure, (b) Computational cost (CPU hours), and (c) Cumulative number of accepted states, all of them as a function of MC steps for both MCDFT and a-MCDFT methods on $6 \times 6 \times 6$ supercells.} }
    \label{fig_aMCDFT_predictions_666}
\end{figure}

%%%-------------------------------------------------
%%%-------------------------\input{5_Conclusions}
%%%-------------------------------------------------

%\section{Conclusions}\label{sec_Conclusions}

In summary, we propose here a novel computational approach to accelerate the discovery of MECs, a critical preliminary step in investigating SRO in MPEAs.
The method, named a-MCDFT, integrates MC sampling, first-principles DFT calculations, and machine learning techniques. 
Grounded in the principles of CE, it incorporates a LOF model that facilitates the rapid and accurate prediction of the energy of the swapped configurations, thereby enhancing the computational efficiency of the traditional MC-DFT framework. 
We validate our methodology through a tungsten-based quaternary and equiatomic HEA case study, but its applicability can be extended to other MPEAs.
The results highlight the accuracy of the a-MCDFT workflow in identifying MECs and calculating their energies, achieving relative errors of just $\sim 0.022\%$ when compared to DFT energy evaluations. 
Furthermore, it is capable of finding such MECs at a fraction of the computational cost of traditional MC-DFT calculations, enabling a much larger number of MC swap trials per atom. 
This success highlights the potential of our computational approach to accurately and efficiently investigate larger supercells \revision{(preliminary results are shown in Fig.~\ref{fig_aMCDFT_predictions_666} and \ref{sect_app_B})}, a critical step towards unraveling the impact of local chemical ordering in the vast structural and chemical space of MPEAs.

%%%-------------------------------------------------
%%%-------------------------
%%%-------------------------------------------------

%\ref{sec:sample:appendix}.

%% The Appendices part is started with the command \appendix;
%% appendix sections are then done as normal sections

\section*{Acknowledgments}
All authors acknowledge support from the U.S. Department of Energy, Office of Science, Fusion Energy Sciences Program Early Career Research Program under Award Number DE-SC0023072. 
%\revision{The authors thank Dr. Jie Peng for useful suggestions and motivating discussions.}
%
The authors acknowledge computer time allocations at Villanova's Augie cluster.
\revision{This research made use of Idaho National Laboratory’s High Performance Computing systems located at the Collaborative Computing Center and supported by the Office of Nuclear Energy of the U.S. Department of Energy and the Nuclear Science User Facilities under Contract No. DE-AC07-05ID14517.}
\revision{This work received funding from Villanova University’s Falvey Memorial Library Scholarship Open Access Reserve (SOAR) Fund.}

%
%This work also used the Advanced Cyberinfrastructure Coordination Ecosystem: Services \& Support (ACCESS) program, which is supported by National Science Foundation grants 2138259, 2138286, 2138307, 2137603, and 2138296. Specifically, the authors acknowledge support from allocation MAT200015.

\newpage
\setcounter{page}{1} % restart page number for Supplemental Materials
\noindent \textbf{Supporting Information}

\begin{center}
Accelerating the discovery of low-energy structure configurations: a computational approach that integrates first-principles calculations, Monte Carlo sampling, and Machine Learning 

\vspace{7mm}

Md Rajib Khan Musa, Yichen Qian, Jie Peng, David Cereceda 

\vspace{5mm}

\footnotesize{Department of Mechanical Engineering, Villanova University, Villanova, 19085, PA, USA} \\

\end{center}

%%%%-----Appendix before references
\appendix

%\section{Sample Appendix Section}
\vspace{-5mm}

\section{Hyperparameter grid search}\label{sect_app_grid_search}

The four hyperparameters of the a-MCDFT method are the cutoff radius (pair and triplet) of the cluster vectors, the size of the training set containing $N_{train}$ pairs of structures with their energies $\left(E_i, S_i \right)$ obtained exclusively via MC-DFT, and the number of nearest neighbors $N_{neighbors}$ needed for the LOF model.
As part of the training process, we performed a grid search over these four key hyperparameters to determine their optimal values that minimize both the error in the energy predictions and the computational cost. 
The grid values for each of the hyperparameters are:  \revision{$N_{train}$} 50:50:500, \revision{$N_{neighbors}$} [5:5:40], pair cutoff [0.5:1.0:10.5], and triple cutoff [0.5:1.0:8.5].
\revision{The computational cost of performing such hyperparameter grid search was $\sim$640 CPU hours.}
When the model is trained with the first $N_{train}$ MC steps of a traditional MC-DFT run, the performance metrics shown in Figure \ref{fig_grid_search} are always calculated against the last 500 MC steps of that same run. 
If the LOF identifies one of the testing configurations as an outlier, that configuration is added to the training set, and the model is retrained before it evaluates the following configuration in the test set. However, the error when predicting the energy of the outlier configuration with the newly trained model (which now includes that configuration and its energy obtained from DFT) is not included in the performance metric.

%%%% ? In Fig.2, if we train for 300 MC steps, calculate the performance metric (R^2, RMSE, MAE) on the last 500 MC steps of the MC-DFT.

The statistical analysis results of such grid search are summarized in Fig.~\ref{fig_grid_search}. As a way of example, Table~\ref{table_hopt} shows the optimal values of the four hyperparameters for each training size when trying to minimize RMSE. Those correspond to the values of the data points shown in Fig.~\ref{fig_grid_search}b.
As we discussed in the manuscript, our goal is to minimize both the error in the energy predictions and the computational cost that increases when increasing the training size. Thus, we chose a training size of 300 pairs of structures with their energies $\left(E_i, S_i \right)$ obtained exclusively via MC-DFT, as they offer the best compromise between RMSE values and computational cost.

\begin{table}[hbtp!]
    \centering
    \caption{Optimal value of the hyperparameters for each training size when RMSE is considered the performance metric.}
    \vspace{2mm}
    \resizebox{\textwidth}{!}{
        \begin{tabular}{ccccc}
            %\hline
            \textbf{Training size} & \textbf{\# neighbors} & \textbf{Pair cutoff (\AA)} & \textbf{Triple cutoff (\AA)} & \textbf{RMSE (meV/atom)} \\
            \hline
            50   & 15 & 7.5 & 5.5 & 1.0022 \\
            100  & 15 & 7.5 & 6.5 & 0.8221 \\
            150  & 15 & 8.5 & 5.5 & 0.7355 \\
            200  & 20 & 8.5 & 0.5 & 0.7143 \\
            250  & 15 & 8.5 & 0.5 & 0.7215 \\
            300  & 25 & 8.5 & 4.5 & 0.6798 \\
            350  & 15 & 7.5 & 4.5 & 0.6943 \\
            400  & 25 & 8.5 & 4.5 & 0.656   \\
            450  & 20 & 8.5 & 4.5 & 0.6389 \\
            500  & 10 & 7.5 & 4.5 & 0.6481 \\
            \hline
        \end{tabular}
    }
    \label{table_hopt}
\end{table}

%With just 50 training data, our model reached an RMSE value of $\sim 1.00 meV/atom$. then with increasing training data RMSE decreased as expected and came to saturation. Pearson correlation value $R^2$ gets saturated after training size 300. From this we decided to fix the training size at 300 for subsequent steps.

% \subsection{Hyper-parameter search algorithm}
% %%------ Hyper-parameter search algorithm 
% \begin{algorithm}[H]
% \caption{Hyper-parameter search}\label{alg_HyperParamSearch}
% \begin{algorithmic}[1]
% \For{$N_{train} \in \{50, 100, 150, \ldots, 500\}$}
%     \State $TrainData = Data[0:N_{train}]$
%     \For{$Pair_{cutoff} \in \{0.5, 1.5, 2.5, \ldots, 9.5\}$}
%         \For{$Triplet_{cutoff} \in \{0.5, 1.5, 2.5, \ldots, 9.5\}$}
%             \State Train LOF model
%             \For{$i = 1, 2, \ldots, N_{test}$}
%                 \If{LOF detects outlier}
%                     \State Append structure to $TrainData$
%                 \EndIf
%             \EndFor
%             \State Train regression model using updated $TrainData$
%             \For{$i = 1, 2, \ldots, N_{test}$}
%                 \State Predict energy $E_{pred}$
%             \EndFor
%             \State Calculate RMSE, $R^2$, and MAE using $E_{pred}$ and $E_{dft}$
%         \EndFor
%     \EndFor
% \EndFor
% \State Output: Optimal hyper-parameters with corresponding evaluation metrics.
% \end{algorithmic}
% \end{algorithm}

%%----------------------------

%\vspace{2cm}

\vspace{-5mm}
\revision{\section{Surrogate model training}\label{sect_app_ML_training}}
We trained the surrogate model using ridge regression, a regularized linear regression method that minimizes the sum of squared residuals while adding a penalty term to prevent over-fitting \cite{hoerl1970ridge, hilt1977ridge}. The objective function minimized in ridge regression is defined as:  

\begin{equation}
\text{Q} =  \left\|  \mathbf{y} - \mathbf{X} \mathbf{w} \right\|^2_2 + \alpha \left\|  \mathbf{w} \right\|^2_2
\end{equation}

 %\text{Minimize: } \| \mathbf{y} - \mathbf{X} \mathbf{w} \|_2^2 + \alpha \| \mathbf{w} \|_2^2

where $\mathbf{y}$ is the target vector, $\mathbf{X}$ is the feature matrix, $\mathbf{w}$ is the vector of coefficients, and $\alpha$ is the regularization parameter. The $\alpha$ hyperparameter controls the trade-off between minimizing the residual sum of squares and the penalty term ($\| \mathbf{w} \|_2^2$), which reduces the magnitude of the coefficients, thereby improving model generalization.

To optimize the $\alpha$ hyperparameter, we used 5-fold cross-validation (CV). In 5-fold CV, the data is split into five equal parts (folds). The model is trained on four folds and validated on the remaining fold, iteratively rotating through all five folds. The process ensures that each data point is used both for training and validation, providing a robust estimate of model performance. The average performance across the folds is used to select the best value of $\alpha$. 
This approach helps balance bias and variance by tuning the regularization strength. A higher $\alpha$ value increases the penalty, leading to simpler models with potentially higher bias but lower variance. Conversely, a lower $\alpha$ value results in more complex models that may overfit the training data.

As previously mentioned, we selected a training size of 300 to balance the trade-off between computational cost and accuracy. This ensured that the model training process remained efficient while retaining sufficient data to capture the underlying patterns effectively. 
\revision{The computational cost of training the surrogate model on the $6 \times 6 \times 6$ supercells was $\sim$1.72 CPU hours.} \\

%%% A natural question might arise: why did we use the cross-validation (CV) method to determine the alpha parameter of the ridge regression model, but relied on grid search for the hyperparameters training_size, pair_cutoff, triplet_cutoff, and #neighbours? The reason is that the first three parameters influence the data extraction process from the atomic trajectory, which is subsequently preprocessed into feature vectors. As a result, they cannot be directly controlled within the training loop and, therefore, cannot be included in the CV process. Additionally, the final parameter, #neighbours, is not part of the surrogate model but instead a hyperparameter of the LOF model.part of this surrogate model, it's a hyperparameter of LOF model.

\section{Energy predictions of MECs}\label{sect_app_B}

Tables~\ref{table_dE_accuracy} \revision{and \ref{table_dE_CPU}} display the accuracy of the a-MCDFT results compared to the baseline DFT method. For the MECs found by a-MCDFT, a DFT calculation was performed to assess the accuracy of the energy predictions made by a-MCDFT. The results show that the errors are very small: \\

\vspace{-6mm}

\begin{table}[h!]
\caption{Accuracy of the a-MCDFT predicted energy when compared with those obtained via DFT calculations for the same MECs\revision{, on $4 \times 4 \times 4$ supercells}. Here, $\delta_e$ is the relative difference between the energies predicted by the a-MCDFT and MC-DFT methods, defined as  $\delta_e = \nicefrac{(E_{a-MCDFT} - E_{DFT})}{E_{DFT}}$.}
\centering
\begin{tabular}{lcccc}
Method       & $E_{\text{a-MCDFT}}$ (eV) & $E_{\text{DFT}}$ (eV) &  $\delta_e$ (\%) & \revision{$\Delta E$ (eV/atom)} \\ \hline
a-MCDFT-1   & -1484.26   & -1483.93    & 0.022  & -0.068      \\ \hline
a-MCDFT-2   & -1484.273  & -1483.93    & 0.023  & -0.065        \\ \hline
a-MCDFT-3   & -1484.132  & -1483.78    & 0.023  & -0.067       \\ \hline
\end{tabular}
\label{table_dE_accuracy}
\end{table}

\vspace{-5mm}

\begin{table}[hbtp!]
\caption{Comparison of computational cost and energy levels achieved by different methods \revision{on $4 \times 4 \times 4$ supercells}. \revision{The values are calculated from 300 MC steps until the end of the MC-DFT or a-MCDFT simulations.}}
\centering
\begin{tabular}{lcc}
Method               & Delta CPU-Hour (${\times 10^5}$) & $\revision{\Delta} E$ (eV) \\ \hline
DFT-1 (300-1k)     & 1.085   & -2.172  \\ \hline
DFT-2 (300-1k)     & 1.041   & -2.408  \\ \hline
a-MCDFT-1 (300-10k) & 0.047   & -3.98  \\ \hline
a-MCDFT-2 (300-10k) & 0.054   & -4.00  \\ \hline
a-MCDFT-3 (300-10k) & 0.047   & -3.85  \\ \hline
\end{tabular}
\label{table_dE_CPU}
\end{table}

\revision{Table~\ref{table_dE_accuracy_666} shows the accuracy of the a-MCDFT results compared to the baseline DFT method on larger $6 \times 6 \times 6$ supercells. To test the scalability of the proposed a-MCDFT method, a-MCDFT simulations on the $6 \times 6 \times 6$ supercells used the same hyperparameters derived from the grid search of the $4 \times 4 \times 4$ supercells. With that same set of hyperparameters, the surrogate model was re-trained again during the first 300 MC steps of the MC-DFT simulations on the $6 \times 6 \times 6$ supercells, and new values of both the regression coefficients and the regularization parameter were calculated before the a-MCDFT method started predicting the energy of new configurations. }

\revision{The comparison of Tables~\ref{table_dE_accuracy} and ~\ref{table_dE_accuracy_666} reveals a significantly larger delta for the $6 \times 6 \times 6$ supercells. While it is true that the presented $6 \times 6 \times 6$ simulations used the same grid search hyperparameters optimized for the $4 \times 4 \times 4$ supercells, which could affect their performance, our ongoing and future efforts are directed toward improving the scalability of the a-MCDFT method for larger supercells and different alloying systems.}

\begin{figure}[hbtp!]
    \centering
    %%trim={<left> <lower> <right> <upper>}
    \includegraphics[width=0.5\textwidth,trim={2cm 0cm 0cm 2.2cm}]{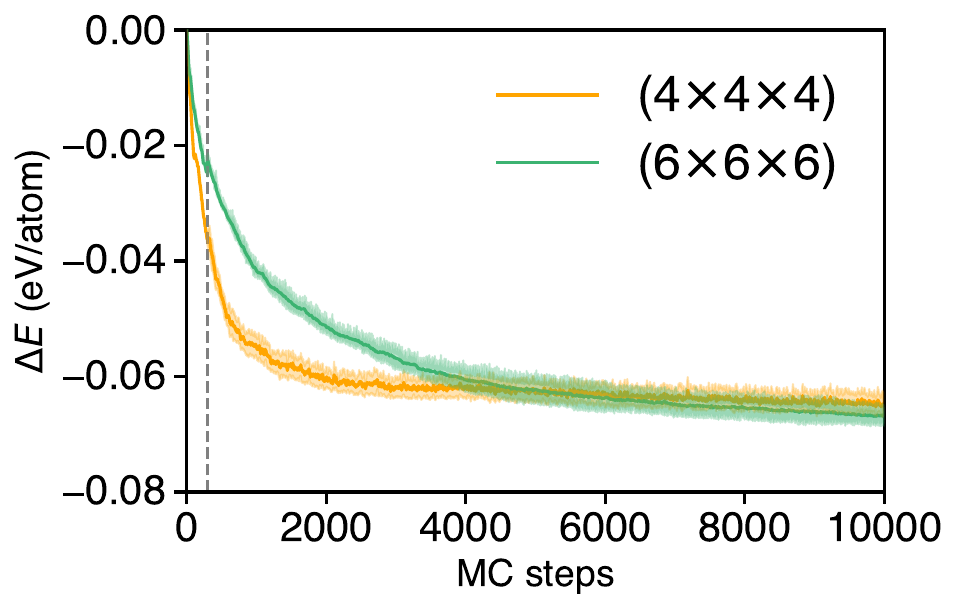}
    %\vspace{-5mm}
    \caption{\revision{Energy evolution per atom (with respect to the energy of the initial SQS structure) predicted by the a-MCDFT approach on both $4 \times 4 \times 4$ and $6 \times 6 \times 6$ supercells. The simulations on the  $6 \times 6 \times 6$ supercells used the same hyperparameters obtained from the grid search on the $4 \times 4 \times 4$ supercells. During the first 300 MC steps, the surrogate models is trained at each supercell size to obtain the most accurate values of the regression coefficients and the regularization parameter.}}
    \label{fig_aMCDFT_444_vs_666}
\end{figure}

\revision{Fig.~\ref{fig_aMCDFT_444_vs_666} shows the energy evolution per atom on both $4 \times 4 \times 4$ and $6 \times 6 \times 6$ supercells.}

\vspace{-4mm}

\begin{table}[h!]
\caption{\revision{Accuracy of the a-MCDFT predicted energy when compared with those obtained via DFT calculations for the same MECs, on $6 \times 6 \times 6$ supercells. To test the scalability of the proposed a-MCDFT method to larger supercells, a-MCDFT simulations on the $6 \times 6 \times 6$ supercells used the same hyperparameters derived from the grid search of the $4 \times 4 \times 4$ supercells.  Here, $\delta_e$ is the relative difference between the energies predicted by the a-MCDFT and MC-DFT methods, defined as  $\delta_e = \nicefrac{(E_{a-MCDFT} - E_{DFT})}{E_{DFT}}$.}}
\centering
\begin{tabular}{lcccc}
Method       & $E_{\text{a-MCDFT}}$ (eV) & $E_{\text{DFT}}$ (eV) &  $\delta_e$ (\%) & $\Delta E$ (eV/atom) \\ \hline
a-MCDFT-1   &   -4557.87     &     -4555.43        &  0.054    &  -0.069  \\ \hline
a-MCDFT-2   &   -4557.00     &     -4554.60        &  0.053    &  -0.067  \\ \hline
a-MCDFT-3   &   -4557.01     &     -4553.40      &  0.08    &   -0.067 \\ \hline
\end{tabular}
\label{table_dE_accuracy_666}
\end{table} 

\vspace{2mm}

\vspace{5mm}

\newpage

\setcounter{page}{13} % continue with page number for AI statement and References

\noindent Statement: During the preparation of this work, the author(s) used Copilot and Grammarly to improve readability and language. After using this tool/service, the author(s) reviewed and edited the content as needed and take(s) full responsibility for the content of the publication.

\newpage

%% If you have bibdatabase file and want bibtex to generate the
%% bibitems, please use
%%

\bibliographystyle{elsarticle-num} 
\bibliography{references_musa1}

\begin{thebibliography}{10}
\expandafter\ifx\csname url\endcsname\relax
  \def\url#1{\texttt{#1}}\fi
\expandafter\ifx\csname urlprefix\endcsname\relax\def\urlprefix{URL }\fi
\expandafter\ifx\csname href\endcsname\relax
  \def\href#1#2{#2} \def\path#1{#1}\fi

\bibitem{yeh2004nanostructured}
J.-W. Yeh, S.-K. Chen, S.-J. Lin, J.-Y. Gan, T.-S. Chin, T.-T. Shun, C.-H.
  Tsau, S.-Y. Chang, Nanostructured high-entropy alloys with multiple principal
  elements: novel alloy design concepts and outcomes, Advanced engineering
  materials 6~(5) (2004) 299--303.

\bibitem{tsai2014high}
M.-H. Tsai, J.-W. Yeh, High-entropy alloys: a critical review, Materials
  Research Letters 2~(3) (2014) 107--123.

\bibitem{li2016metastable}
Z.~Li, K.~G. Pradeep, Y.~Deng, D.~Raabe, C.~C. Tasan, Metastable high-entropy
  dual-phase alloys overcome the strength--ductility trade-off, Nature
  534~(7606) (2016) 227--230.

\bibitem{george2019high}
E.~P. George, D.~Raabe, R.~O. Ritchie, High-entropy alloys, Nature reviews
  materials 4~(8) (2019) 515--534.

\bibitem{murty2019high}
B.~S. Murty, J.-W. Yeh, S.~Ranganathan, P.~Bhattacharjee, High-entropy alloys,
  Elsevier, 2019.

\bibitem{han2024multifunctional}
L.~Han, S.~Zhu, Z.~Rao, C.~Scheu, D.~Ponge, A.~Ludwig, H.~Zhang, O.~Gutfleisch,
  H.~Hahn, Z.~Li, et~al., Multifunctional high-entropy materials, Nature
  Reviews Materials (2024) 1--20.

\bibitem{li2019strengthening}
Q.-J. Li, H.~Sheng, E.~Ma, Strengthening in multi-principal element alloys with
  local-chemical-order roughened dislocation pathways, Nature communications
  10~(1) (2019) 3563.

\bibitem{ding2018tunable}
J.~Ding, Q.~Yu, M.~Asta, R.~O. Ritchie, Tunable stacking fault energies by
  tailoring local chemical order in crconi medium-entropy alloys, Proceedings
  of the National Academy of Sciences 115~(36) (2018) 8919--8924.

\bibitem{zhang2020short}
R.~Zhang, S.~Zhao, J.~Ding, Y.~Chong, T.~Jia, C.~Ophus, M.~Asta, R.~O. Ritchie,
  A.~M. Minor, Short-range order and its impact on the crconi medium-entropy
  alloy, Nature 581~(7808) (2020) 283--287.

\bibitem{chen2021direct}
X.~Chen, Q.~Wang, Z.~Cheng, M.~Zhu, H.~Zhou, P.~Jiang, L.~Zhou, Q.~Xue,
  F.~Yuan, J.~Zhu, et~al., Direct observation of chemical short-range order in
  a medium-entropy alloy, Nature 592~(7856) (2021) 712--716.

\bibitem{hsiao2022data}
H.-W. Hsiao, R.~Feng, H.~Ni, K.~An, J.~D. Poplawsky, P.~K. Liaw, J.-M. Zuo,
  Data-driven electron-diffraction approach reveals local short-range ordering
  in crconi with ordering effects, Nature communications 13~(1) (2022) 6651.

\bibitem{smith2024competition}
N.~C. Smith, T.-c. Liu, Y.~Xia, C.~Wolverton, Competition between long-and
  short-range order in size-mismatched medium-entropy alloys, Acta Materialia
  277 (2024) 120199.

\bibitem{miracle2017critical}
D.~B. Miracle, O.~N. Senkov, A critical review of high entropy alloys and
  related concepts, Acta materialia 122 (2017) 448--511.

\bibitem{han2024ubiquitous}
Y.~Han, H.~Chen, Y.~Sun, J.~Liu, S.~Wei, B.~Xie, Z.~Zhang, Y.~Zhu, M.~Li,
  J.~Yang, et~al., Ubiquitous short-range order in multi-principal element
  alloys, Nature Communications 15~(1) (2024) 6486.

\bibitem{zhang2021improving}
Y.~Zhang, D.~Han, X.~Li, Improving the stress-controlled fatigue life of low
  solid-solution hardening ni-cr alloys by enhancing short range ordering
  degree, International Journal of Fatigue 149 (2021) 106266.

\bibitem{feng2017effects}
W.~Feng, Y.~Qi, S.~Wang, Effects of short-range order on the magnetic and
  mechanical properties of feconi (alsi) x high entropy alloys, Metals 7~(11)
  (2017) 482.

\bibitem{walsh2021magnetically}
F.~Walsh, M.~Asta, R.~O. Ritchie, Magnetically driven short-range order can
  explain anomalous measurements in crconi, Proceedings of the National Academy
  of Sciences 118~(13) (2021) e2020540118.

\bibitem{liu2018effect}
M.~Liu, A.~Aiello, Y.~Xie, K.~Sieradzki, The effect of short-range order on
  passivation of fe-cr alloys, Journal of The Electrochemical Society 165~(11)
  (2018) C830.

\bibitem{xie2021percolation}
Y.~Xie, D.~M. Artymowicz, P.~P. Lopes, A.~Aiello, D.~Wang, J.~L. Hart,
  E.~Anber, M.~L. Taheri, H.~Zhuang, R.~C. Newman, et~al., A percolation theory
  for designing corrosion-resistant alloys, Nature materials 20~(6) (2021)
  789--793.

\bibitem{scully2020controlling}
J.~R. Scully, S.~B. Inman, A.~Y. Gerard, C.~D. Taylor, W.~Windl, D.~K.
  Schreiber, P.~Lu, J.~E. Saal, G.~S. Frankel, Controlling the corrosion
  resistance of multi-principal element alloys, Scripta Materialia 188 (2020)
  96--101.

\bibitem{blades2024tuning}
W.~H. Blades, B.~Redemann, N.~Smith, D.~Sur, M.~Barbieri, Y.~Xie, S.~Lech,
  E.~Anber, M.~Taheri, C.~Wolverton, et~al., Tuning chemical short-range order
  for stainless behavior at reduced chromium concentrations in multi-principal
  element alloys, arXiv preprint arXiv:2403.00086 (2024).

\bibitem{zhang2023effect}
Z.~Zhang, Z.~Su, B.~Zhang, Q.~Yu, J.~Ding, T.~Shi, C.~Lu, R.~O. Ritchie, E.~Ma,
  Effect of local chemical order on the irradiation-induced defect evolution in
  crconi medium-entropy alloy, Proceedings of the National Academy of Sciences
  120~(15) (2023) e2218673120.

\bibitem{lu2017radiation}
C.~Lu, T.~Yang, K.~Jin, N.~Gao, P.~Xiu, Y.~Zhang, F.~Gao, H.~Bei, W.~J. Weber,
  K.~Sun, et~al., Radiation-induced segregation on defect clusters in
  single-phase concentrated solid-solution alloys, Acta Materialia 127 (2017)
  98--107.

\bibitem{cao2021does}
P.~Cao, How does short-range order impact defect kinetics in irradiated
  multiprincipal element alloys?, Accounts of Materials Research 2~(2) (2021)
  71--74.

\bibitem{zhang2017local}
F.~Zhang, S.~Zhao, K.~Jin, H.~Xue, G.~Velisa, H.~Bei, R.~Huang, J.~Ko,
  D.~Pagan, J.~Neuefeind, et~al., Local structure and short-range order in a
  nicocr solid solution alloy, Physical review letters 118~(20) (2017) 205501.

\bibitem{joress2023exafs}
H.~Joress, B.~Ravel, E.~Anber, J.~Hollenbach, D.~Sur, J.~Hattrick-Simpers,
  M.~L. Taheri, B.~DeCost, Why is exafs for complex concentrated alloys so
  hard? challenges and opportunities for measuring ordering with x-ray
  absorption spectroscopy, Matter (2023).

\bibitem{hastings1970monte}
W.~K. Hastings, Monte carlo sampling methods using markov chains and their
  applications, Biometrika (1970).

\bibitem{tamm2015atomic}
A.~Tamm, A.~Aabloo, M.~Klintenberg, M.~Stocks, A.~Caro, Atomic-scale properties
  of ni-based fcc ternary, and quaternary alloys, Acta Materialia 99 (2015)
  307--312.

\bibitem{kang2023density}
S.~Kang, A.~Tamm, Density functional study of atomic arrangements in crmnfeconi
  high-entropy alloy and their impact on vacancy formation energy and
  segregation, Computational Materials Science 230 (2023) 112456.

\bibitem{samanta2024machine}
A.~Samanta, P.~Balaprakash, S.~Aubry, B.~K. Lin, Machine-learning-aided density
  functional theory calculations of stacking fault energies in steel, Scripta
  Materialia 241 (2024) 115862.

\bibitem{sheriff2024quantifying}
K.~Sheriff, Y.~Cao, T.~Smidt, R.~Freitas, Quantifying chemical short-range
  order in metallic alloys, Proceedings of the National Academy of Sciences
  121~(25) (2024) e2322962121.

\bibitem{cao2024capturing}
Y.~Cao, K.~Sheriff, R.~Freitas, Capturing short-range order in high-entropy
  alloys with machine learning potentials, arXiv preprint arXiv:2401.06622
  (2024).

\bibitem{yichen_mcdft_2024}
Y.~Qian, M.~R. Gilbert, L.~Dezerald, D.~Nguyen-Manh, D.~Cereceda,
  First-principles study of the energetics and the local chemical ordering of
  tungsten-based alloys, arXiv preprint arXiv:2410.03998 (2024).

\bibitem{Kikuchi1951}
R.~Kikuchi, \href{https://doi.org/10.1103/PhysRev.81.988}{A theory of
  cooperative phenomena}, Physical Review 81~(6) (1951) 988.
\newblock \href {https://doi.org/10.1103/PhysRev.81.988}
  {\path{doi:10.1103/PhysRev.81.988}}.
\newline\urlprefix\url{https://doi.org/10.1103/PhysRev.81.988}

\bibitem{Sanchez1984}
J.~M. Sanchez, F.~Ducastelle, D.~Gratias,
  \href{https://doi.org/10.1016/0378-4371(84)90096-7}{Generalized cluster
  description of multicomponent systems}, Physica A: Statistical Mechanics and
  its Applications 128~(1-2) (1984) 334--350.
\newblock \href {https://doi.org/10.1016/0378-4371(84)90096-7}
  {\path{doi:10.1016/0378-4371(84)90096-7}}.
\newline\urlprefix\url{https://doi.org/10.1016/0378-4371(84)90096-7}

\bibitem{Zunger2002}
A.~Zunger, S.~Wagner, A.~Franceschetti, C.~Wolverton, S.~H. Wei,
  \href{https://doi.org/10.1088/0965-0393/10/6/304}{Obtaining ising-like
  expansions for binary alloys from first principles}, Modelling and Simulation
  in Materials Science and Engineering 10~(6) (2002) 685.
\newblock \href {https://doi.org/10.1088/0965-0393/10/6/304}
  {\path{doi:10.1088/0965-0393/10/6/304}}.
\newline\urlprefix\url{https://doi.org/10.1088/0965-0393/10/6/304}

\bibitem{DeFontaine1994}
D.~De~Fontaine, \href{https://doi.org/10.1016/S0081-1947(08)60040-5}{Cluster
  approach to order-disorder transformations in alloys}, Solid State Physics 47
  (1994) 33--176.
\newline\urlprefix\url{https://doi.org/10.1016/S0081-1947(08)60040-5}

\bibitem{breunig2000lof}
M.~M. Breunig, H.-P. Kriegel, R.~T. Ng, J.~Sander, Lof: Identifying
  density-based local outliers, ACM SIGMOD Record 29~(2) (2000) 93--104.

\bibitem{sobieraj2020chemical}
D.~Sobieraj, J.~S. Wr{\'o}bel, T.~Rygier, K.~J. Kurzyd{\l}owski, O.~El~Atwani,
  A.~Devaraj, E.~M. Saez, D.~Nguyen-Manh, Chemical short-range order in
  derivative cr--ta--ti--v--w high entropy alloys from the first-principles
  thermodynamic study, Physical Chemistry Chemical Physics 22~(41) (2020)
  23929--23951.

\bibitem{el2019outstanding}
O.~El-Atwani, N.~Li, M.~Li, A.~Devaraj, J.~Baldwin, M.~M. Schneider,
  D.~Sobieraj, J.~S. Wr{\'o}bel, D.~Nguyen-Manh, S.~A. Maloy, et~al.,
  Outstanding radiation resistance of tungsten-based high-entropy alloys,
  Science advances 5~(3) (2019) eaav2002.

\bibitem{waseem2017powder}
O.~A. Waseem, H.~J. Ryu, Powder metallurgy processing of a wxtativcr
  high-entropy alloy and its derivative alloys for fusion material
  applications, Scientific reports 7~(1) (2017) 1926.

\bibitem{zou2015ultrastrong}
Y.~Zou, H.~Ma, R.~Spolenak, Ultrastrong ductile and stable high-entropy alloys
  at small scales, Nature communications 6~(1) (2015) 1--8.

\bibitem{senkov2018development}
O.~N. Senkov, D.~B. Miracle, K.~J. Chaput, J.-P. Couzinie, Development and
  exploration of refractory high entropy alloys—a review, Journal of
  materials research 33~(19) (2018) 3092--3128.

\bibitem{el2021helium}
O.~El-Atwani, A.~Alvarado, K.~Unal, S.~Fensin, J.~Hinks, G.~Greaves,
  J.~Baldwin, S.~Maloy, E.~Martinez, Helium implantation damage resistance in
  nanocrystalline w-ta-v-cr high entropy alloys, Materials Today Energy 19
  (2021) 100599.

\bibitem{senkov2011mechanical}
O.~N. Senkov, G.~B. Wilks, J.~M. Scott, D.~B. Miracle, Mechanical properties of
  nb25mo25ta25w25 and v20nb20mo20ta20w20 refractory high entropy alloys,
  Intermetallics 19~(5) (2011) 698--706.

\bibitem{qian2021using}
Y.~Qian, M.~R. Gilbert, L.~Dezerald, D.~Cereceda, Using first-principles
  calculations to predict the mechanical properties of transmuting tungsten
  under first wall fusion power-plant conditions, Journal of Physics: Condensed
  Matter 33~(34) (2021) 345901.

\bibitem{qian2023ab}
Y.~Qian, M.~R. Gilbert, L.~Dezerald, D.~Nguyen-Manh, D.~Cereceda, Ab initio
  study of tungsten-based alloys under fusion power-plant conditions, Journal
  of Nuclear Materials (2023) 154422.

\bibitem{kresse1993ab}
G.~Kresse, J.~Hafner, Ab initio molecular dynamics for liquid metals, Physical
  review B 47~(1) (1993) 558.

\bibitem{blochl1994projector}
P.~E. Bl{\"o}chl, Projector augmented-wave method, Physical review B 50~(24)
  (1994) 17953.

\bibitem{PBE}
J.~P. Perdew, K.~Burke, M.~Ernzerhof, Generalized gradient approximation made
  simple, Physical review letters 77~(18) (1996) 3865.

\bibitem{handbook1992alloy}
A.~Handbook, Alloy phase diagrams, vol 03. asm international, materials park, p
  556 and 1107 (1992).

\bibitem{hoerl1970ridge}
A.~E. Hoerl, R.~W. Kennard, Ridge regression: Biased estimation for
  nonorthogonal problems, Technometrics 12~(1) (1970) 55--67.

\bibitem{hilt1977ridge}
D.~E. Hilt, D.~W. Seegrist, Ridge, a computer program for calculating ridge
  regression estimates, Department of Agriculture, Forest Service, Northeastern
  Forest Experiment~…, 1977.

\end{thebibliography}

%% else use the following coding to input the bibitems directly in the
%% TeX file.

% \begin{thebibliography}{00}

% %% \bibitem{label}
% %% Text of bibliographic item

% \bibitem{}

% \end{thebibliography}

%%%%-----Appendix after references so it can be taken as separate file for Supplementary Materials (Scripta Materialia)
\newpage

% \appendix

% %\section{Sample Appendix Section}
% \input{Appendix_A} 

% \newpage

% \input{Appendix_B} 

\end{document}